\documentclass[aps,prl,twocolumn,floatfix,groupedaddress]{revtex4-2}
\usepackage{graphicx,amsmath,amsfonts,mathrsfs,bbold,dsfont}
\begin{document}

\title{Molecular-orbital origin of weak orbital ferromagnetism and anomalous Hall effect in a trimerized kagome antiferromagnet}

\author{I.~V.~Solovyev \url{(SOLOVYEV.Igor@nims.go.jp)}}
\affiliation{Research Center for Materials Nanoarchitectonics (MANA), National Institute for Materials Science (NIMS), 1-1 Namiki, Tsukuba, Ibaraki 305-0044, Japan}

\date{\today}

\date{\today}

\maketitle

\begin{widetext}
\section*{ABSTRACT}
\par We reveal the molecular-orbital origin of weak orbital ferromagnetism and the anomalous Hall effect (AHE) induced by noncollinear spin order in a trimerized kagome lattice. By transforming to the molecular-orbital basis of an isolated trimer in the double-exchange regime, we derive an effective low-energy model for the states with opposite orbital angular momenta $m= \pm 1$ subject to an effective Zeeman field arising from spin chirality and spin-orbit (SO) coupling. In this representation, the kagome-lattice problem maps onto a triangular lattice of trimers, while the nearest- and next-nearest-neighbor hoppings of the original lattice generate complex hopping amplitudes between neighboring trimers. This formulation provides a transparent picture of how spin chirality and SO coupling can be recast as an effective Zeeman field that lifts the degeneracy of the $m= \pm 1$ states, breaks time-reversal symmetry, and thereby gives rise to weak orbital ferromagnetism and the AHE. The resulting phase diagram includes metallic phases, topologically trivial insulating phases, as well as several topological phases characterized by nonzero Chern numbers and the quantum anomalous Hall effect. We further identify an alternative route to time-reversal-symmetry breaking and orbital magnetization, driven by an emergent gauge field arising solely from complex inter-trimer hoppings and requiring no effective Zeeman field.
\end{widetext}

\section*{INTRODUCTION}
\par The emergence of time-reversal-odd electronic responses in antiferromagnetic (AFM) materials has attracted considerable attention owing to both their fundamental significance and their potential for spintronic applications. Compared with conventional ferromagnets, antiferromagnets produce negligible stray magnetic fields, are less susceptible to external magnetic perturbations, and can support spin dynamics at terahertz frequencies~\cite{Jungwirth,Baltz}. Despite their vanishing net spin magnetization, antiferromagnets can exhibit finite orbital magnetization and a variety of time-reversal-odd electronic responses, including the anomalous Hall effect (AHE) and magneto-optical effects~\cite{TurovShavrov,PRB1997}, that are commonly regarded as hallmarks of ferromagnetism.

\par Current research on time-reversal-odd electronic responses in antiferromagnets largely centers on two broad classes of materials: nearly collinear and noncollinear antiferromagnets~\cite{SmejkalNRM}. The study of such phenomena in nearly collinear antiferromagnets has a long history. A particularly important class is formed by centrosymmetric materials with antipolar distortions, where the doubled crystallographic unit cell accommodates the antiferromagnetic order without further enlargement~\cite{Dzyaloshinskii1991,PRB2026}. In addition to weak spin ferromagnetism~\cite{BorovikRomanov,Dzyaloshinskii_weakF}, such systems can exhibit a variety of other `ferromagnetic' properties~\cite{TurovBook,TurovUFN}, including the AHE, which was predicted already in 1963~\cite{TurovShavrov}. The recent resurgence of interest in these materials has been stimulated by the concept of altermagnetism~\cite{SmejkalPRX1}, characterized by the emergence of momentum-dependent non-relativistic spin splitting in compensated collinear antiferromagnets, as noted already in 1981~\cite{Sandratskii1981}. 

\par Nevertheless, altermagnetic band splitting is not the sole mechanism underlying time-reversal-odd electronic responses in nearly collinear antiferromagnets. For example, the AHE in centrosymmetric antiferromagnets is driven by relativistic spin-orbit (SO) interactions, while the contribution of the non-relativistic band splitting itself may be negligibly small~\cite{NakaOrganic,Naka_Spintronics,npjQM2026}. Another important aspect is that, in centrosymmetric antiferromagnets with antipolar distortions, the relativistic SO interaction transforms in the same way as the N\'eel order parameter. As a consequence, their electronic structure can be described using a generalized Bloch theorem and mapped onto that of an effective ferromagnet in an appropriately chosen local coordinate frame~\cite{PRB2026}. Although the AHE and weak spin ferromagnetism originate from different microscopic mechanisms~\cite{NakaOrganic,npjQM2026}, the AHE is intrinsically linked to the emergence of a net orbital magnetization~\cite{PRB2026,npjQM2026}. From this perspective, the AHE can be regarded as a manifestation of weak orbital ferromagnetism. Furthermore, in the local coordinate frame the staggered N\'eel exchange field is transformed into a uniform Zeeman field acting on the electronic degrees of freedom. This observation provides a particularly transparent connection to ferromagnetic systems~\cite{PRB2026}.

\par The situation is quite different in noncollinear antiferromagnets realized on triangular~\cite{Shindou,CoNb3S6Ghimire,Seki,CoNb3S6Khanh} and kagome~\cite{Ohgushi,Kubler,ChenNiuMacDonald,SCZhang2,Nakatsuji,Nakatsuji2,Felser,Felser2,Arita,Guo,Mokrousov,Mertig} lattices. Here, the crystallographic unit cell likewise contains several magnetic atoms but, in contrast to centrosymmetric collinear antiferromagnets, their number is typically odd. Consequently, the magnetic structure cannot be reduced to two spin sublattices related by spin reversal. Instead, the magnetic order is naturally described by three or more spin sublattices forming a noncollinear texture within the crystallographic unit cell. In these systems, the microscopic origin of the AHE depends crucially on the magnetic texture. For coplanar spin textures, such as the $120^{\circ}$ spin structure in the kagome plane, the effect requires relativistic SO interactions~\cite{ChenNiuMacDonald,Kubler,Arita,Mertig}. In noncoplanar spin textures, by contrast, the AHE can arise from a finite scalar spin chirality $\chi = \boldsymbol{e}_{1} \cdot [\boldsymbol{e}_{2} \times \boldsymbol{e}_{3}]$, associated with the relative orientation of three spins on a triangle, even in the absence of SO coupling~\cite{Ohgushi,Shindou,Arita,Kubler}. Thus, $\chi$ acts as an emergent magnetic field, giving rise to the so-called topological AHE~\cite{Bruno} and net orbital magnetization~\cite{Lounis}.

\par In this work, we show that despite these differences, antipolarly distorted centrosymmetric antiferromagnets and noncollinear antiferromagnets on the kagome lattice share common microscopic principles of time-reversal-symmetry breaking and the emergence of orbital ferromagnetism. To reveal this connection, we reformulate the kagome-lattice problem in the molecular basis that diagonalizes the electronic Hamiltonian of an isolated magnetic trimer in the double-exchange regime. In this representation, the kagome lattice is mapped onto a triangular lattice whose elementary unit is the trimer itself. In the strong-trimerization limit, the low-energy physics is described by an effective $2\times 2$ Hamiltonian, where both scalar spin chirality and SO coupling enter through an effective Zeeman field that splits the molecular-orbital states with angular momenta $m = \pm 1$. The resulting effective model establishes a direct parallel between weak orbital ferromagnetism in centrosymmetric antiferromagnets and kagome-lattice noncollinear antiferromagnets. At the same time, it reveals a rich phase diagram comprising metallic phases, topologically trivial insulating phases, and several topological phases characterized by nonzero Chern numbers, whose existence was anticipated in earlier studies of kagome systems~\cite{Ohgushi,SCZhang2}. Finally, we demonstrate that time-reversal symmetry can be broken and a finite orbital magnetization can be induced even in the absence of the effective Zeeman field through an appropriate rescaling of the complex hopping parameters, highlighting the broader range of mechanisms available in trimerized kagome systems.

\par Although the model analysis is rather general, we closely follow the specific example of Mn$_3$Ge, which exhibits a giant AHE according to both experimental observations~\cite{Nakatsuji2} and electronic structure calculations~\cite{Kubler,Felser2,Arita,Guo}.

\section*{Results}
\subsection*{Notations and basic symmetry properties} 
\par
\begin{figure}[b]
\begin{center}
\includegraphics[width=8.6cm]{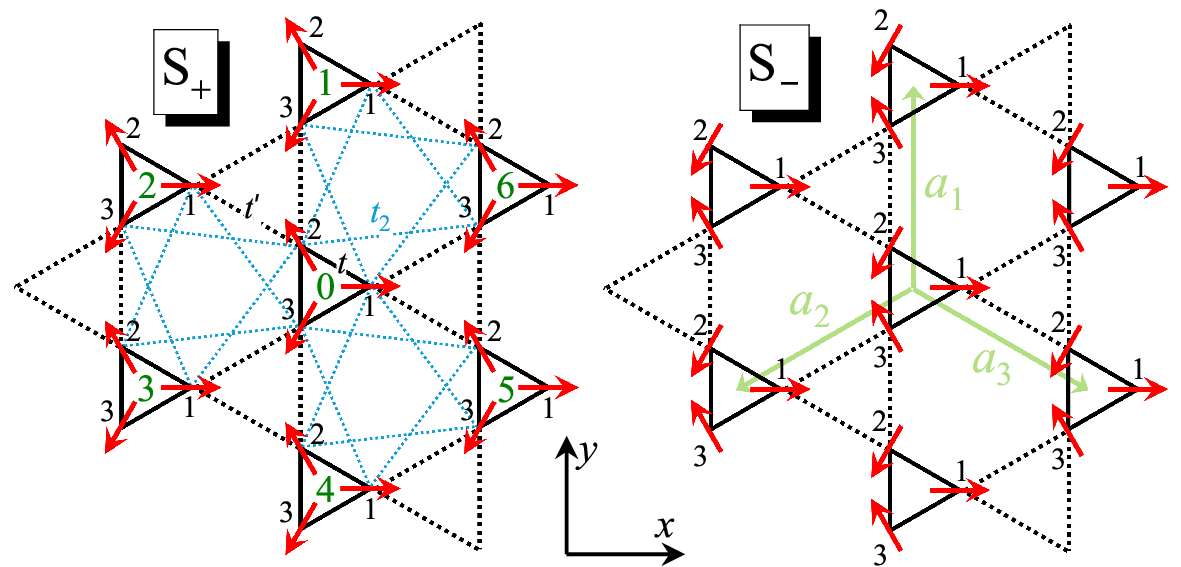} 
\end{center}
\caption{Two possible spin configurations on the kagome plane, $\mathrm{S}_{+}$ and $\mathrm{S}_{-}$. Red arrows indicate the projections of the spins onto the $xy$ plane. Green arrows show the triangular-lattice translation vectors for the $\mathrm{S}_{-}$ configuration, while the hopping parameters are labeled for the $\mathrm{S}_{+}$ configuration. Thick solid bonds denote the intra-trimer hopping $t$, whereas black and blue dashed bonds represent the nearest-neighbor and next-nearest-neighbor inter-trimer hoppings $t'$ and $t_{2}$, respectively. Large olive symbols label the trimers, and small black symbols label the three sites within each trimer.}
\label{fig:kagome}
\end{figure}

 We consider two noncoplanar spin configurations on the kagome lattice with opposite scalar chirality, $\chi_{\pm} = \pm \frac{3\sqrt{3}}{2}\sin^2 \vartheta \cos \vartheta$, which are denoted by $\mathrm{S}_{+}$ and $\mathrm{S}_{-}$. For the $\mathrm{S}_{+}$ state, the spin directions on the three sublattices are 
\noindent
\begin{eqnarray}
\boldsymbol{e}_{1} & = & (\sin \vartheta, 0, \cos \vartheta ), \\
\boldsymbol{e}_{2} & = & (-\frac{1}{2} \sin \vartheta, \frac{\sqrt{3}}{2} \sin \vartheta, \cos \vartheta ), \\ 
\boldsymbol{e}_{3} & = & (-\frac{1}{2} \sin \vartheta,-\frac{\sqrt{3}}{2} \sin \vartheta, \cos \vartheta ). 
\end{eqnarray}
\noindent The $\mathrm{S}_{-}$ state is obtained by interchanging sublattices $2$ and $3$, while $\boldsymbol{e}_{1}$ remains unchanged, as illustrated in Fig.~\ref{fig:kagome}. With this convention, $\mathrm{S}_{+}$ and $\mathrm{S}_{-}$ correspond to positive and negative scalar chirality, respectively.
\noindent

\par Throughout this work, $\mathcal{R}$ denotes rotations of the spin magnetic moments, whereas $\mathcal{C}$ refers to rotations of the lattice. Since time reversal ($\mathcal{T}$) flips all spin directions, $\mathcal{T}\boldsymbol{e}_{n}=-\boldsymbol{e}_{n}$, the two chiral states are related by $\mathcal{R}_{2y}\mathcal{T}\mathrm{S}_{+} = \mathrm{S}_{-}$, where $\mathcal{R}_{2y}$ represents a $180^{\circ}$ rotation of the spin vector about the $y$ axis. Furthermore, both $\mathrm{S}_{+}$ and $\mathrm{S}_{-}$ are invariant under the combined threefold rotation $\mathcal{R}_{3z}\mathcal{C}_{3z}$, which simultaneously rotates the lattice and the spins about the $z$ axis.

\par The special case $\vartheta = \frac{\pi}{2}$ corresponds to a coplanar magnetic structure with vanishing scalar chirality. In this limit, the spins form the conventional $120^{\circ}$ kagome texture. The configurations $\mathrm{S}_{+}$ and $\mathrm{S}_{-}$ differ only by the sense of spin rotation around an elementary triangle, corresponding to $+120^{\circ}$ and $-120^{\circ}$ (or equivalently $240^{\circ}$) rotations, respectively. They can be transformed into one another either by the spin rotation $\mathcal{R}_{2x}$ or by the lattice rotation $\mathcal{C}_{2x}$. Consequently, each coplanar state is invariant under the combined operation $\mathcal{R}_{2x}\mathcal{C}_{2x}$.

\par Unless stated otherwise, we focus on the $\mathrm{S}_{+}$ state throughout the remainder of this paper. 

\par For the untrimerized kagome lattice, spatial inversion about the center of a hexagon is a symmetry operation. The trimerization, however, breaks inversion symmetry. As a result, physical phenomena that are forbidden in the centrosymmetric lattice may become allowed in the trimerized phase, which plays an important role in the electronic and transport properties discussed below.

\subsection*{Model} 
\par We consider a one-orbital model on a trimerized kagome lattice. The model includes the nearest-neighbor (nn) intra-trimer hopping $t$, the nn inter-trimer hopping $t'$, and the next-nearest-neighbor (nnn) inter-trimer hopping $t_{2}$, as illustrated in Fig.~\ref{fig:kagome}. The nnn hopping $t_{2}$ is essential for realizing time-reversal-symmetry breaking in the absence of an explicit Zeeman field.

\subsubsection{\label{sec:DE} Double exchange regime}
\par The anomalous Hall effect in kagome-lattice systems is often discussed in the double-exchange (DE) regime~\cite{Ohgushi}, where the exchange splitting between majority- and minority-spin states is sufficiently large that the minority-spin sector can be projected out. This projection is performed in a local coordinate frame whose quantization $z$ axis is aligned with $\boldsymbol{e}_{n}$. As a result, the problem can be formulated entirely in the basis of majority-spin states~\cite{deGennes}. 

\par More specifically, let $\hat{U}_{n}$ denote the SU(2) rotation matrix that aligns the local spin quantization axis with $\boldsymbol{e}_{n}$. In the DE limit, an operator $\hat{O}_{nn'}$ is transformed to the local spin frame according to $\hat{O}_{nn'} \to \hat{U}_{n}^{\phantom{\dagger}} \hat{O}_{nn'} \hat{U}_{n'}^{\dagger}$, after which only the majority-spin component, corresponding to the upper-left matrix element, is retained.

\par This procedure renormalizes the hopping amplitudes by a complex factor $A e^{\pm i\Phi}$ with
\noindent
\begin{equation}
A = \frac{\sqrt{3\cos^2\vartheta + 1}}{2}
\end{equation}
\noindent and
\begin{equation}
\Phi = \pm \arctan \left( \frac{\sqrt{3}(1-\cos \vartheta)}{1 + 3\cos \vartheta} \right). 
\end{equation}
\noindent The sign in $A e^{\pm i\Phi}$ depends on the bond orientation: the positive (negative) sign applies to the $1$-$2$ ($1$-$3$) and to all bonds related to it by even permutations of the site labels $1$, $2$, and $3$. The two spin configurations differ only by the sign of $\Phi$: $+\Phi$ for $\mathrm{S}_{+}$ and $-\Phi$ for $\mathrm{S}_{-}$. The coplanar configuration corresponds to $\Phi = \pm \frac{\pi}{3}$. 

\subsubsection{\label{sec:SO} Spin-orbit interaction}
\par In the one-orbital model, the SO coupling is introduced by adding the spin-dependent hopping term $\hat{\mathcal{H}}^{\rm so}_{nn'} = i\boldsymbol{\lambda}_{nn'}\cdot \hat{\boldsymbol{\sigma}}$ defined on the noncentrosymmetric bonds, where $\hat{\boldsymbol{\sigma}}$ denotes the vector of spin Pauli matrices. The total hopping amplitude is therefore $\hat{t}_{nn'}=t + i\boldsymbol{\lambda}_{nn'}\cdot \hat{\boldsymbol{\sigma}}$. The axial vectors $\boldsymbol{\lambda}_{nn'}$ obey the same symmetry constraints as the Dzyaloshinskii-Moriya (DM) interactions~\cite{Dzyaloshinskii_weakF,Moriya_weakF}.

\par
\begin{figure}[b]
\begin{center}
\includegraphics[width=8.6cm]{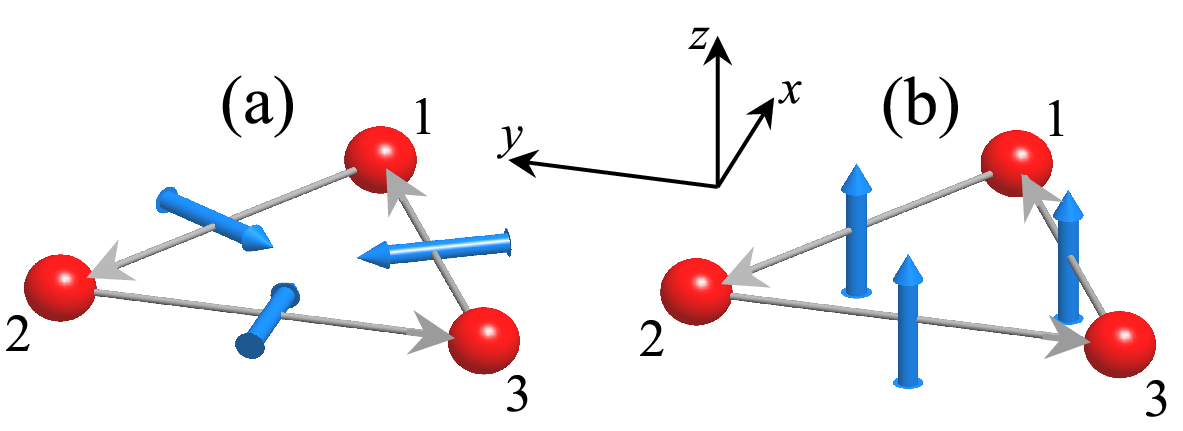} 
\end{center}
\caption{Possible alignments of the spin-orbit interaction vectors allowed by $\mathcal{C}_{3z}$ symmetry: (a) within the trimers and (b) perpendicular to the trimers. Bond directions are indicated by grey arrows.}
\label{fig:SO}
\end{figure}

 Figs.~\ref{fig:SO}(a) and (b) show two possible patterns of the SO vectors $\boldsymbol{\lambda}_{nn'}$ consistent with the $\mathcal{C}_{3z}$ symmetry of the kagome lattice. For the in-plane configuration shown in Fig.~\ref{fig:SO}(a), the SO vectors are $\boldsymbol{\lambda}_{12} = (-\frac{1}{2},-\frac{\sqrt{3}}{2},0)\lambda^{\perp}$, $\boldsymbol{\lambda}_{23} = (1,0,0)\lambda^{\perp}$, and $\boldsymbol{\lambda}_{31} = (-\frac{1}{2},\frac{\sqrt{3}}{2},0)\lambda^{\perp}$. They lie in the $xy$ plane and are perpendicular to the corresponding bonds. For the out-of-plane configuration shown in Fig.~\ref{fig:SO}(b), all SO vectors are oriented along the $z$ axis, $\boldsymbol{\lambda}_{nn'}=(0,0,\lambda^{\parallel})$.
\noindent

\par The SO contribution to the effective DE hopping can be written as $\lambda A_{\mathrm{so}} e^{\pm i\Phi_{\mathrm{so}}}$, with the signs corresponding to the same bond directions as for the spinless hoppings. For the in-plane configuration of the SO vectors, $\lambda = \lambda^{\perp}$, $A_{\mathrm{so}}^{\perp} = \sin \vartheta$, and $\Phi_{\mathrm{so}}^{\perp} = \mp \frac{\pi}{6}$. For the out-of-plane configuration, $\lambda = \lambda^{\parallel}$,
\noindent
\begin{equation}
A_{\mathrm{so}}^{\parallel} = \frac{\sqrt{3+\cos^2\vartheta}}{2},  
\end{equation}
\noindent and
\noindent
\begin{equation}
\Phi_{\mathrm{so}}^{\parallel} = \pm \arctan \left( \frac{3+\cos \vartheta}{\sqrt{3}(1 - \cos \vartheta)} \right). 
\end{equation}

\par The AFM exchange coupling tends to stabilize the coplanar configuration, while the DM interactions lift the degeneracy between the $\mathrm{S}_{+}$ and $\mathrm{S}_{-}$ states. Which state is realized as the ground state depends on the sign of $\lambda^{\perp}$ or $\lambda^{\parallel}$ (Supplementary, Sec.~A). 

\par In Mn$_3$Ge, the symmetry allowed intra-trimer SO vectors are $\boldsymbol{\lambda}_{nn'} = (0,0,\lambda^{\parallel})$, i.e., perpendicular to the kagome plane. Effective in-plane components arise only through inter-trimer processes that combine inter-plane hopping with SO coupling (Supplementary, Sec.~B). This naturally explains why the Hall conductivity $\sigma_{zx}$, which directly depends on the inter-plane SO coupling, is considerably larger than $\sigma_{xy}$, as suggested by experiment~\cite{Nakatsuji2} and electronic-structure calculations~\cite{Felser2}. Despite this microscopic constraint, phenomenological models of kagome-lattice systems are often supplemented by an in-plane SO coupling term~\cite{ChenNiuMacDonald,Mertig}.

\subsubsection{\label{sec:trimer} Single trimer}
\par The Hamiltonian for a single trimer is given by
\noindent
\begin{equation}
\hat{\mathcal{H}}_{\bigtriangleup} = At
\left(
\begin{array}{ccc}
0 & e^{i\Phi} & e^{-i\Phi} \\
e^{-i\Phi} & 0 & e^{i\Phi} \\
e^{i\Phi} & e^{-i\Phi} & 0 
\end{array}
\right).
\label{eq:Hsingle}
\end{equation}
\noindent Introducing the gauge transformation $\hat{R} = \textrm{diag}(0,\omega,\omega^2)$, $\omega = e^{\frac{2 \pi i}{3}}$, one readily verifies that $\hat{R} \hat{\mathcal{H}}_{\bigtriangleup}(\Phi) \hat{R}^{\dagger} = \hat{\mathcal{H}}_{\bigtriangleup}(\Phi-\frac{2\pi}{3})$. In particular, for the coplanar configuration ($\Phi=\frac{\pi}{3}$), $\hat{R} \hat{\mathcal{H}}_{\bigtriangleup}(\frac{\pi}{3}) \hat{R}^{\dagger} = \hat{\mathcal{H}}_{\bigtriangleup}(-\frac{\pi}{3})=\hat{\mathcal{H}}_{\bigtriangleup}^{*}(\frac{\pi}{3})$. Thus, the action of time reversal (complex conjugation) can be compensated by a gauge transformation, implying that $\hat{\mathcal{H}}_{\bigtriangleup}(\frac{\pi}{3})$ is time-reversal invariant up to a gauge transformation.

\par The eigenvalues of $\hat{\mathcal{H}}_{\bigtriangleup}$ are $\varepsilon_{k} = 2 At \cos \left( \frac{2\pi k}{3} + \Phi \right)$, where $k=$ $0$, $1$, $2$ denote the discrete crystal momenta of the three-site trimer. The corresponding normalized eigenvectors are independent of $\Phi$ and can be written as
\noindent
\begin{equation}
| k \rangle = \frac{1}{\sqrt{3}}\left(
\begin{array}{c}
1 \\
\omega^{k} \\
\omega^{2k}
\end{array}
\right).
\end{equation}
\noindent Introducing the matrix,
\noindent
\begin{equation}
\hat{D} = \frac{1}{\sqrt{3}}\left(
\begin{array}{ccc}
1 & 1 & 1 \\
1 & \omega^2 & \omega \\
1 & \omega & \omega^2
\end{array}
\right),
\end{equation}
\noindent the Hamiltonian is transformed to the diagonal representation $\hat{\tilde{\mathcal{H}}}_{\bigtriangleup} = \hat{D} \hat{\mathcal{H}}_{\bigtriangleup} \hat{D}^{\dagger} = \textrm{diag}(\varepsilon_{0},\varepsilon_{1},\varepsilon_{2})$.

\par The matrix representation of the orbital magnetic moment operator $\hat{L}^{z}$, constrained by the $\mathcal{C}_{3z}$ symmetry and the requirement that it be odd under time reversal (complex conjugation), $\hat{L}^{z} \xrightarrow{\mathcal{T}} - \hat{L}^{z}$, is
\noindent
\begin{displaymath}
\hat{L}^{z} = \left(
\begin{array}{ccc}
 0 & i\mu & -i\mu \\
 -i\mu & 0 & i\mu \\
i\mu & -i\mu & 0
\end{array}
\right).
\end{displaymath}
\noindent This operator acts only in the orbital sector and is therefore independent of the spin indices. The parameter $i\mu$ denotes the corresponding angular-momentum matrix element in the Wannier basis. In the DE subspace, the projected orbital angular momentum operator is $\hat{\mathcal{L}}^{z} = \frac{\mu}{t}\frac{\partial \hat{\mathcal{H}}_{\bigtriangleup}}{\partial \Phi}$. Consequently, $\hat{\mathcal{L}}^{z}$ commutes with $\hat{\mathcal{H}}_{\bigtriangleup}$, and the trimer eigenstates $| k \rangle$ remain simultaneous eigenstates of both operators. The corresponding orbital moments are $\langle k | \hat{\mathcal{L}}^{z} | k \rangle = -2A\mu \sin \left( \frac{2\pi k}{3} + \Phi \right)$. 

\par
\begin{figure}[b]
\begin{center}
\includegraphics[width=8.6cm]{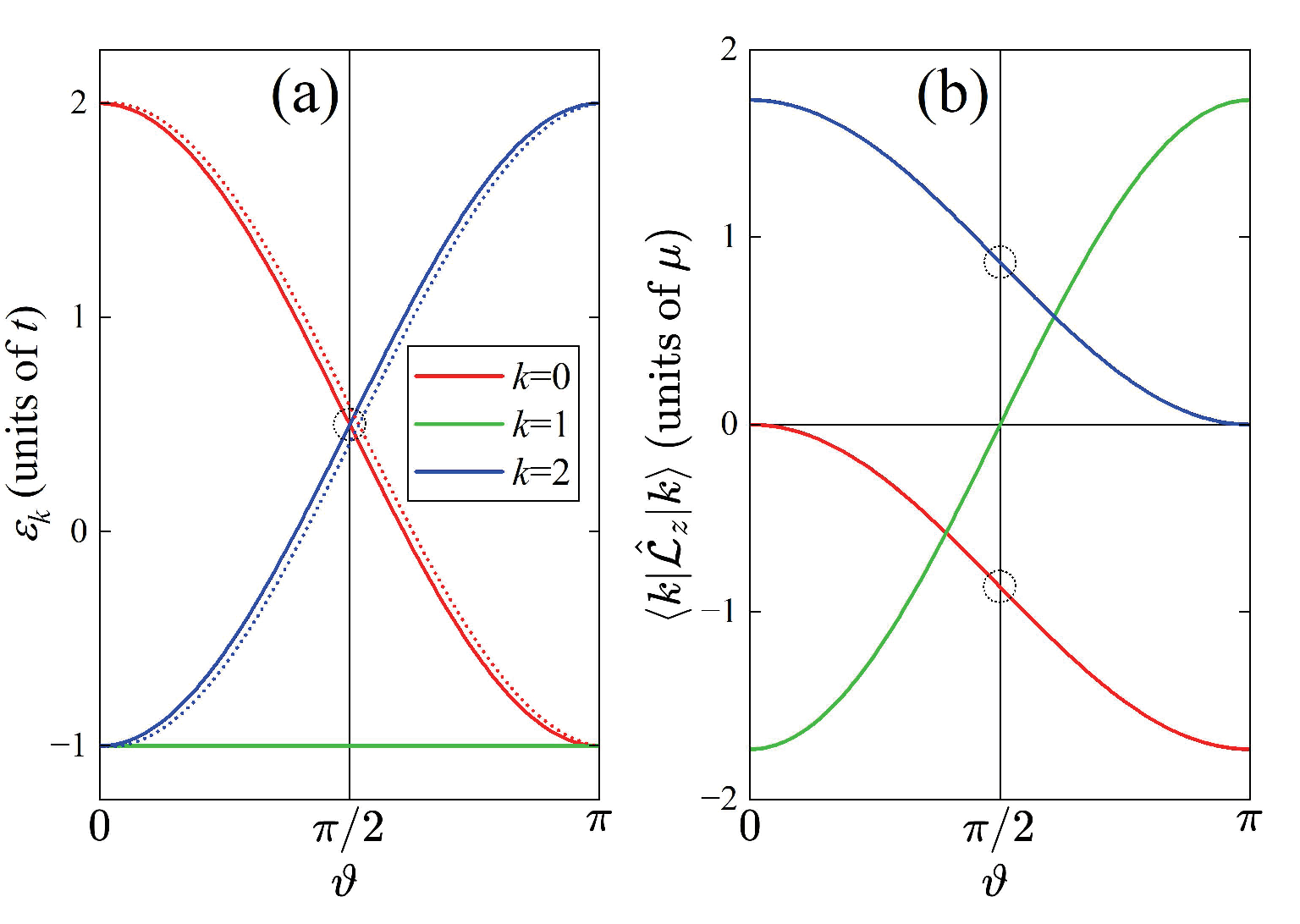} 
\end{center}
\caption{Properties of a single trimer as a function of the polar angle $\vartheta$ for the $\textrm{S}_{+}$ state: (a) energy spectrum with (dotted line) and without (solid line) in-plane spin-orbit coupling, $\lambda^{\perp} = 0.1t$; (b) expectation values of the orbital angular-momentum. The states corresponding to the coplanar spin configuration at $\vartheta = \frac{\pi}{2}$ are highlighted by circles.}
\label{fig:trimer}
\end{figure}

 The results for the $\textrm{S}_{+}$ state are summarized in Fig.~\ref{fig:trimer}. For the coplanar spin configuration at $\vartheta = \frac{\pi}{2}$, corresponding to $\Phi = \frac{\pi}{3}$, the states with $k=0$ and $k=2$ become degenerate. These states are separated from the $k=1$ state by an energy gap of $\frac{3|t|}{2}$. Therefore, in the limit of strong trimerization, where $|t|$ greatly exceeds the inter-trimer hoppings $t'$ and $t_2$, the high-energy $k=1$ state can be projected out. The low-energy sector is then described by a $2 \times 2$ Hamiltonian in the subspace spanned by the $k=0$ and $k=2$ states with opposite orbital magnetic momenta $\langle k | \hat{\mathcal{L}}^{z} | k \rangle = \mp \frac{\sqrt{3}}{2} \mu$, and can thus be identified with molecular-orbital states carrying angular momenta $m = \pm 1$.
\noindent

\par Mn$_3$Ge appears to lie in the strong-trimerization regime. Indeed, let $d^{S(L)}_{\mathrm{Mn-Mn}}$ denote the short (long) Mn-Mn bond length within (between) the trimers. Then, the experimental distortion corresponds to $\frac{d^{L}_{\mathrm{Mn-Mn}}}{d^{S}_{\mathrm{Mn-Mn}}} = 1.208$~\cite{Mn3Gestr}. Using Harrison's scaling $t(d) \sim d^{-5}$ for hopping amplitudes between Mn $3d$ orbitals~\cite{Harrison}, one obtains estimates $\frac{|t'|}{|t|} \sim 0.39$. This estimate places Mn$_3$Ge on the trimerized side of the crossover and supports the use of the strong-trimerization framework as a starting point.

\par Trimerization of the kagome lattice is known to qualitatively alter its low-energy electronic structure and can promote the formation of molecular-orbital states localized on triangular clusters~\cite{Damski,Bolens,npjQM2021}. In the present work, however, its primary role is to simplify the electronic structure by removing the $m=0$ states from the low-energy sector, thereby isolating the physics associated with the nearly degenerate $m= \pm 1$ states.

\par Having established the relevant low-energy doublet, we now examine how its degeneracy is lifted away from the coplanar limit. Writing $\Phi$ as $\Phi = \frac{\pi}{3} - \Delta \Phi$, one finds that for $\vartheta$ close to $\frac{\pi}{2}$, $\Delta \Phi \approx \sqrt{3} \left( \frac{\pi}{2} - \vartheta \right) = \frac{2}{3}\chi_{+}$. Expanding the trimer energies around $\Phi = \frac{\pi}{3}$ yields $\varepsilon_{0} \approx \frac{t}{2} \left( 1 + \sqrt{3} \Delta \Phi \right)$ and $\varepsilon_{2} \approx \frac{t}{2} \left( 1 - \sqrt{3} \Delta \Phi \right)$. The resulting splitting of the ($k=0$, $k=2$) doublet is therefore characterized by the effective field
\noindent
\begin{equation}
B=\frac{\varepsilon_{0}-\varepsilon_{2}}{2} = \frac{\sqrt{3}t \Delta \Phi}{2} = \frac{t\chi_{+}}{\sqrt{3}}.
\label{eq:B}
\end{equation}
\noindent The scalar spin chirality therefore acts as an effective orbital Zeeman field that lifts the degeneracy of the trimer doublet. In the subspace spanned by $|0 \rangle$ and $|2 \rangle$, the corresponding contribution to the effective Hamiltonian is $B\hat{\tau}_{z}$, where $\hat{\tau}_{z}$ acts on the orbital pseudospin degree of freedom. 

\par The SO interaction provides an additional mechanism for splitting the $k=0$ and $k=2$ trimer levels. Since the SO term has the same eigenstates as $\hat{\mathcal{H}}_{\bigtriangleup}$, its effect is controlled by the relative phase $\Phi_{\mathrm{so}}-\Phi$. For the out-of-plane configuration of the SO vectors [Fig.~\ref{fig:SO}(b)] and the coplanar spin texture ($\vartheta = \frac{\pi}{2}$), one has $\Phi_{\mathrm{so}} = \Phi = \frac{\pi}{3}$. In this case, the SO interaction merely renormalizes the hopping amplitude, $t \to t + \sqrt{3} \lambda^{\parallel}$ and therefore leaves the $k=0$ and $k=2$ states degenerate. By contrast, for the in-plane configuration [Fig.~\ref{fig:SO}(a)], $\Phi_{\mathrm{so}} = -\frac{\pi}{6}$, which shifts the condition for the level crossing. To leading order, this shift can be expressed as $\vartheta \to \vartheta - \frac{\lambda^{\parallel}}{3t}$. Since the effective field $B$ is determined solely by $\vartheta$, the shift is equivalent to the renormalization $B \to B + \frac{\lambda^{\parallel}}{2}$. Thus, the SO interaction lifts the degeneracy at $\vartheta=\frac{\pi}{2}$, while an accidental degeneracy reappears at $\vartheta=\frac{\pi}{2} + \frac{\lambda^{\parallel}}{3t}$. The SO coupling therefore shifts the position of the level crossing without changing its underlying structure (Fig.~\ref{fig:trimer}). This behavior was first pointed out in Ref.~\cite{Mertig}; here we provide a transparent microscopic explanation of its origin.

\par Therefore, in the following we limit our discussion to the effective field $B$ arising from spin chirality, bearing in mind that the same formalism also applies to contributions from the in-plane SO coupling.

\par The analysis for the $\textrm{S}_{-}$ state proceeds analogously. The essential difference is that time reversal interchanges the states $|1 \rangle$ and $|2 \rangle$. Consequently, at $\vartheta = \frac{\pi}{2}$ the degeneracy occurs between the states $|1 \rangle$ and $|0 \rangle$, which form the low-energy subspace in the strong-trimerization limit. To maintain a common pseudospin convention for both chiralities, the basis states are ordered according to the sign of their orbital moment: the first basis state is taken to have negative $\langle k | \hat{\mathcal{L}}^{z} | k \rangle$, while the second basis state has positive $\langle k | \hat{\mathcal{L}}^{z} | k \rangle$. Accordingly, the pseudospin basis is chosen as $(|0 \rangle,|2 \rangle)$ for $\textrm{S}_{+}$ and $(|1 \rangle,|0 \rangle)$ for $\textrm{S}_{-}$.

\subsubsection{\label{sec:Hk} Effective Hamiltonian}
\par The inter-trimer hoppings can be expressed in the molecular basis that diagonalizes the single-trimer Hamiltonian $\hat{\mathcal{H}}_{\bigtriangleup}$  (Supplementary, Sec.~C). This representation allows the standard one-orbital model with three atomic sites per kagome unit cell to be reformulated as a three-orbital model with a single molecular site. In this basis, both the nn and nnn inter-trimer hoppings of the atomic model map solely onto nn hoppings between neighboring molecular sites. Consequently, the system can be viewed as an effective model on a triangular lattice with a single molecular site per unit cell and only nn hoppings.

\par Under strong trimerization ($|t| \gg |t'|$, $|t| \gg |t_{2}|$), the molecular orbital with $k=1$ remains well separated in energy and can be integrated out. The electronic Hamiltonian can therefore be restricted to the subspace spanned by the $k=0$ and $k=2$ molecular orbitals. In reciprocal space, this reduced $2 \times 2$ Hamiltonian can be written in the general form:
\noindent
\begin{equation}
\hat{\mathcal{H}}_{\boldsymbol{k}} = \mathrm{h}^{0}_{\boldsymbol{k}} \hat{\mathbb{1}} + \mathbf{h}^{\phantom{0}}_{\boldsymbol{k}} \cdot \hat{\boldsymbol{\tau}} + B \hat{\tau}_{z},
\label{eq:Hk}
\end{equation}
\noindent where $\hat{\mathbb{1}}$ and $\hat{\boldsymbol{\tau}}=(\hat{\tau}_{x},\hat{\tau}_{y},\hat{\tau}_{z})$ denote the identity and Pauli matrices acting in the pseudospin subspace of the $k=0$ and $k=2$ molecular orbitals, while the effective Zeeman field is given by Eq.~(\ref{eq:B}). Both $\mathrm{h}^{0}_{\boldsymbol{k}}$ and $\mathbf{h}^{\phantom{0}}_{\boldsymbol{k}}=(\mathrm{h}^{x}_{\boldsymbol{k}},\mathrm{h}^{y}_{\boldsymbol{k}},\mathrm{h}^{z}_{\boldsymbol{k}})$ receive contributions from the nn and nnn hoppings of the original kagome lattice. Accordingly, we write $\mathrm{h}^{\alpha}_{\boldsymbol{k}} = h^{\alpha}_{\boldsymbol{k}} + \Delta h^{\alpha}_{\boldsymbol{k}}$ ($\alpha=0$, $x$, $y$, $z$), where the first and second terms originate from the nn and nnn hoppings, respectively. These functions are obtained by Fourier transforming the hopping amplitudes on the triangular lattice. The lattice translation vectors are $\boldsymbol{a}_{1}=(0,1)$, $\boldsymbol{a}_{2}=(-\frac{\sqrt{3}}{2},-\frac{1}{2})$, and $\boldsymbol{a}_{3}=(\frac{\sqrt{3}}{2},-\frac{1}{2})$. Defining $\gamma_{i}^{\pm} = \boldsymbol{k}\cdot \boldsymbol{a}_{i} \pm \frac{2\pi}{3}$ and $\gamma_{i}^{0} = \boldsymbol{k}\cdot \boldsymbol{a}_{i}$ ($i=1$, $2$, $3$), the results can be summarized as follows. The nn hopping contributions are: 
\noindent
\begin{eqnarray}
h^{0}_{\boldsymbol{k}} & = & \phantom{-}\frac{t'}{6} \left( \cos \gamma^{0}_{1} + \cos \gamma^{0}_{2} + \cos \gamma^{0}_{3} \right), \label{eq:hk1} \\
h^{x}_{\boldsymbol{k}} & = & -\frac{t'}{6} \left( 2\cos \gamma^{0}_{1} - \cos \gamma^{0}_{2} - \cos \gamma^{0}_{3} \right), \label{eq:hk2} \\
h^{y}_{\boldsymbol{k}} & = & -\frac{t'}{2\sqrt{3}} \left( \cos \gamma^{0}_{2} - \cos \gamma^{0}_{3} \right), \label{eq:hk3} \\
h^{z}_{\boldsymbol{k}} & = & -\frac{t'}{2\sqrt{3}} \left( \sin \gamma^{0}_{1} + \sin \gamma^{0}_{2} + \sin \gamma^{0}_{3} \right). \label{eq:hk4}
\end{eqnarray}
\noindent The nnn hopping contributions are: 
\noindent
\begin{eqnarray}
\Delta h^{0}_{\boldsymbol{k}} & = & \frac{t_{2}}{3} \left( \cos \gamma^{0}_{1} + \cos \gamma^{0}_{2} + \cos \gamma^{0}_{3} \right), \label{eq:dhk1} \\
\Delta h^{x}_{\boldsymbol{k}} & = & \frac{t_{2}}{6} \left( 2\cos \gamma^{0}_{1} - \cos \gamma^{0}_{2} - \cos \gamma^{0}_{3} \right), \label{eq:dhk2} \\
\Delta h^{y}_{\boldsymbol{k}} & = & \frac{t_{2}}{2\sqrt{3}} \left( \cos \gamma^{0}_{2} - \cos \gamma^{0}_{3} \right), \label{eq:dhk3} \\ 
\Delta h^{z}_{\boldsymbol{k}} & = & \frac{t_{2}}{\sqrt{3}} \left( \sin \gamma^{0}_{1} + \sin \gamma^{0}_{2} + \sin \gamma^{0}_{3} \right). \label{eq:dhk4}
\end{eqnarray}
\noindent To leading order, the $\Delta \Phi$-dependence of the hopping amplitudes can be neglected. The resulting corrections are of order $t' \Delta \Phi$, which are parametrically smaller than the Zeeman field $B \sim t \Delta \Phi$ in the strong-trimerization regime.

\par The Hamiltonian (\ref{eq:Hk}) has eigenvalues 
\noindent
\begin{equation}
\varepsilon_{\boldsymbol{k}}^{\pm} = \mathrm{h}^{0}_{\boldsymbol{k}} \pm |\mathbf{h}^{\phantom{0}}_{\boldsymbol{k}}+\boldsymbol{B}|.
\label{eq:ek}
\end{equation}
\noindent The corresponding normalized eigenvectors can be written in the standard spinor form as~\cite{npjQM2026}: 
\noindent
\begin{equation}
| u^{-}_{\boldsymbol{k}} \rangle = \left(
\begin{array}{c}
\cos \theta_{\boldsymbol{k}} e^{i \phi_{\boldsymbol{k}}} \\
\sin \theta_{\boldsymbol{k}}
\end{array}
\right), \quad 
| u^{+}_{\boldsymbol{k}} \rangle = \left(
\begin{array}{c}
-\sin \theta_{\boldsymbol{k}} \\
\cos \theta_{\boldsymbol{k}} e^{-i \phi_{\boldsymbol{k}}}
\end{array}
\right) , 
\label{eq:uk}
\end{equation}
\noindent where $\boldsymbol{B}=(0,0,B)$,
\noindent 
\begin{equation}
\theta_{\boldsymbol{k}}  = \frac{1}{2} \arctan \frac{ \sqrt{( \mathrm{h}^{x}_{\boldsymbol{k}})^{2} + (\mathrm{h}^{y}_{\boldsymbol{k}})^{2}}} {\mathrm{h}^{z}_{\boldsymbol{k}}+B}  \quad (0 \le \theta_{\boldsymbol{k}} < \pi), 
\label{eq:thetak}
\end{equation}
\noindent and
\noindent
\begin{equation}
\phi_{\boldsymbol{k}} = - \arctan \left( \frac{\mathrm{h}^{y}_{\boldsymbol{k}}}{\mathrm{h}^{x}_{\boldsymbol{k}}} \right)  \quad  (0 \le \phi_{\boldsymbol{k}} < 2\pi).
\label{eq:phik}
\end{equation}

\subsubsection{\label{sec:Berry} Berry curvature and anomalous Hall conductivity}
\par The anomalous Hall conductivity is given by the Brillouin‑zone (BZ) integral of the Berry curvature $\Omega^{\nu}_{\boldsymbol{k}} = -2{\rm Im} \left\langle \partial_{k_{x}} u^{\nu}_{\boldsymbol{k}} | \partial_{k_{y}} u^{\nu}_{\boldsymbol{k}} \right\rangle$, weighted by the Fermi–Dirac distribution $f^{\nu}_{\boldsymbol{k}}$~\cite{Fang,Yao,NagaosaRevModPhys}:
\noindent
\begin{displaymath}
\sigma_{xy} =  - \frac{e^2}{\hbar} \sum_{\nu = \pm} \int_{\rm BZ} \frac{d \boldsymbol{k}}{(2 \pi)^{2}} f^{\nu}_{\boldsymbol{k}} \Omega^{\nu}_{\boldsymbol{k}} .
\end{displaymath}
\noindent For a two-band Hamiltonian, the Berry curvature can be expressed in terms of the winding number of $\mathbf{h}^{\phantom{0}}_{\boldsymbol{k}}+\boldsymbol{B}$ around the origin~\cite{SCZhang}:
\noindent
\begin{equation}
\Omega^{-}_{\boldsymbol{k}} =  -\Omega^{+}_{\boldsymbol{k}} = \frac{(\mathbf{h}^{\phantom{0}}_{\boldsymbol{k}}+\boldsymbol{B}) \cdot [\partial_{k_{x}} \mathbf{h}^{\phantom{0}}_{\boldsymbol{k}} \times \partial_{k_{y}} \mathbf{h}^{\phantom{0}}_{\boldsymbol{k}}] }{2|\mathbf{h}^{\phantom{0}}_{\boldsymbol{k}}+\boldsymbol{B}|^3}. 
\label{eq:Berryk}
\end{equation}

\par
\begin{figure}[t]
\begin{center}
\includegraphics[width=8.6cm]{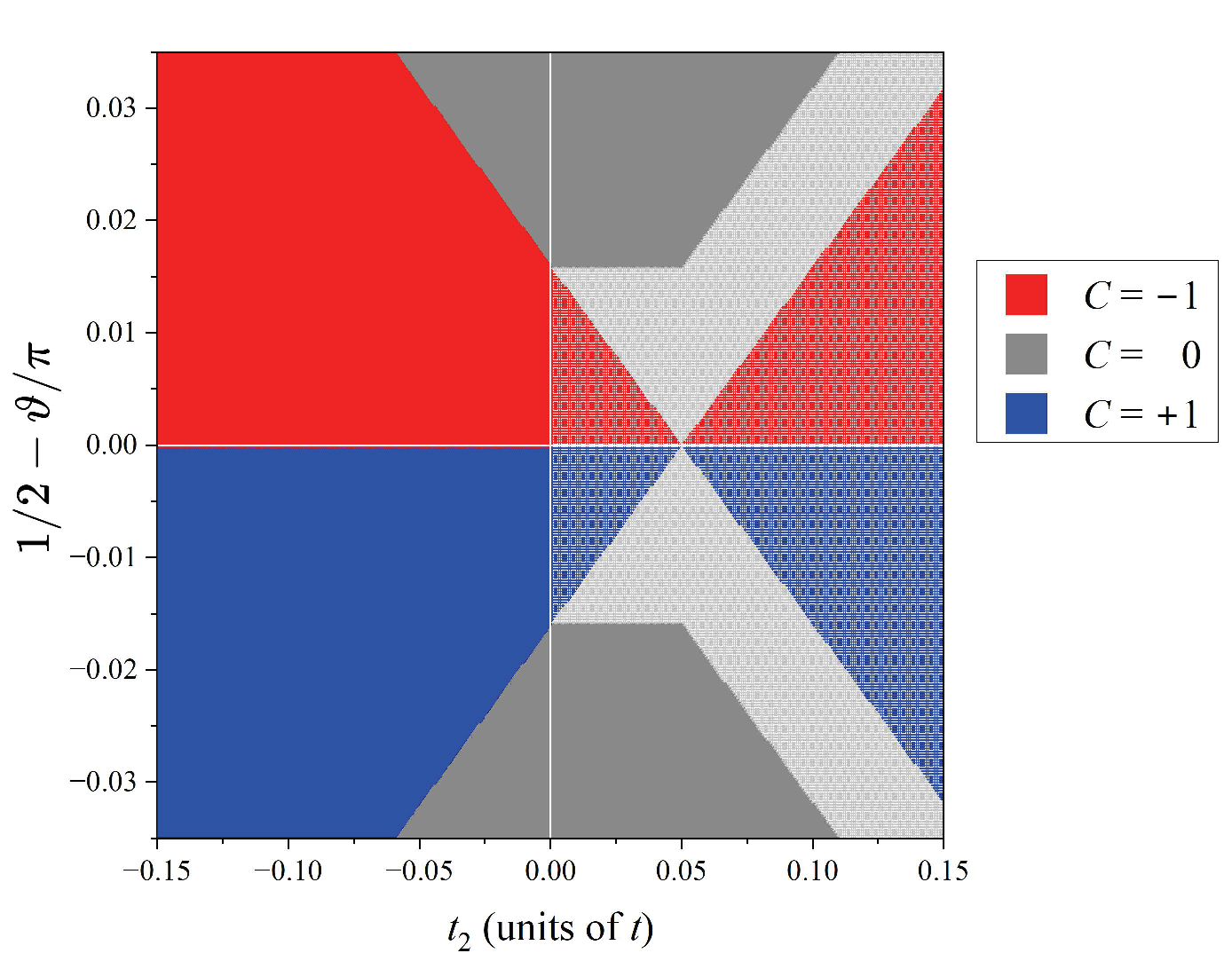} 
\end{center}
\caption{Phase diagram at half filling as a function of the next-nearest-neighbor hopping $t_{2}$ and the polar angle $\vartheta$ shown near the coplanar configuration ($\vartheta=\frac{\pi}{2}$). The nearest-neighbor inter-trimer hopping is fixed at $t' = 0.1t$, where $t$ denotes the intra-trimer hopping. Regions with different Chern numbers are shown in distinct colors, as indicated in the legend. The deep-colored areas correspond to insulating phases with a finite gap separating the two bands, whereas the pale-colored regions indicate intrinsically metallic behavior: the two bands avoid crossing at every $\boldsymbol{k}$-point yet overlap in energy, so that no global gap opens at half filling.}
\label{fig:PhaseD}
\end{figure}

 At half filling, the topology of the insulating state is characterized by the Chern number, defined as
\noindent
\begin{equation}
C = \frac{1}{2\pi} \int_{\rm BZ} d \boldsymbol{k} \Omega^{-}_{\boldsymbol{k}}.
\label{eq:Cherm}
\end{equation}
\noindent In insulating phases, the Hall conductivity is quantized according to $\sigma_{xy} = -\frac{e^2}{h}C$. The resulting phase diagram for $t'=0.1t$ is shown in Fig.~\ref{fig:PhaseD} as a function of the next-nearest-neighbor hopping $t_{2}$ and the polar angle $\vartheta$, which parameterizes the effective field $B$ that splits the $k=0$ and $k=2$ molecular orbitals. 
\noindent

\par Beyond the trivial insulating regimes realized when $\vartheta$ departs significantly from $\frac{\pi}{2}$, the phase diagram also hosts topological phases, including Chern metals and Chern insulators with $C=\pm 1$. 

\subsubsection{\label{sec:TR} Time-reversal symmetry}
\par Using Eqs.~(\ref{eq:hk1})-(\ref{eq:hk4}) and (\ref{eq:dhk1})-(\ref{eq:dhk4}), one readily finds  
\noindent
\begin{equation}
\mathrm{h}^{0}_{-\boldsymbol{k}}=\mathrm{h}^{0}_{\boldsymbol{k}}, \ \mathrm{h}^{x}_{-\boldsymbol{k}}=\mathrm{h}^{x}_{\boldsymbol{k}}, \ \mathrm{h}^{y}_{-\boldsymbol{k}}=\mathrm{h}^{y}_{\boldsymbol{k}}, \  \mathrm{h}^{z}_{-\boldsymbol{k}}=-\mathrm{h}^{z}_{\boldsymbol{k}}.
\label{eq:TRinv}
\end{equation} 
\noindent For $B=0$, Eq.~(\ref{eq:Berryk}) then gives $\Omega^{\nu}_{-\boldsymbol{k}} = -\Omega^{\nu}_{\boldsymbol{k}}$, as expected for systems possessing time-reversal symmetry. At first sight, Eq.~(\ref{eq:TRinv}) appears inconsistent with the conventional time-reversal symmetry of a spinless two-band Hamiltonian. Under complex conjugation $K$, the momentum transforms as $\boldsymbol{k} \to -\boldsymbol{k}$ and the Pauli matrix $\hat{\tau}_{y}$ changes sign because it is purely imaginary. Therefore, the usual spinless time-reversal condition $\hat{\mathcal{H}}_{-\boldsymbol{k}}^{*}=\hat{\mathcal{H}}_{\boldsymbol{k}}$ would require $\mathrm{h}^{0}_{-\boldsymbol{k}}=\mathrm{h}^{0}_{\boldsymbol{k}}$, $\mathrm{h}^{x}_{-\boldsymbol{k}}=\mathrm{h}^{x}_{\boldsymbol{k}}$, $\mathrm{h}^{y}_{-\boldsymbol{k}}=-\mathrm{h}^{y}_{\boldsymbol{k}}$, and $\mathrm{h}^{z}_{-\boldsymbol{k}}=\mathrm{h}^{z}_{\boldsymbol{k}}$. This definition, however, is incomplete because, within the reduced molecular-orbital subspace, time reversal is not represented by complex conjugation alone. Under $K$, the $\mathrm{S}_{+}$ state is transformed into $\mathrm{S}_{-}$, and the molecular orbital $| 2 \rangle$ is mapped to $| 1 \rangle$, while $| 0 \rangle$ remains unchanged. Because the pseudospin basis is defined as ($| 0 \rangle$, $| 2 \rangle$) in the $\mathrm{S}_{+}$ sector but as ($| 1 \rangle$, $| 0 \rangle$) in the $\mathrm{S}_{-}$ sector, time reversal effectively exchanges the two pseudospin components. Therefore, time reversal is not represented by complex conjugation alone within the reduced pseudospin subspace. Since it exchanges the two molecular-orbital states that define the pseudospin basis, its action must be supplemented by $\hat{\tau}_{x}$, which leaves $\hat{\tau}_{x}$ invariant while sending $\hat{\tau}_{y} \to - \hat{\tau}_{y}$ and $\hat{\tau}_{z} \to - \hat{\tau}_{z}$. Thus, the appropriate antiunitary symmetry is $\mathcal{T}=\hat{\tau}_{x}K$ and the corresponding condition for time-reversal invariance is $\hat{\tau}_{x}\hat{\mathcal{H}}_{-\boldsymbol{k}}^{*}\hat{\tau}_{x}=\hat{\mathcal{H}}_{\boldsymbol{k}}$. This condition directly leads to the relations given in Eq.~(\ref{eq:TRinv}).

\par For finite $B$, one has $\varepsilon_{-\boldsymbol{k}}^{\nu}(-B) = \varepsilon_{\boldsymbol{k}}^{\nu}(B)$ and $\Omega_{-\boldsymbol{k}}^{\nu}(-B) = -\Omega_{\boldsymbol{k}}^{\nu}(B)$, as follows from Eqs.~(\ref{eq:ek}), (\ref{eq:Berryk}), and (\ref{eq:TRinv}). Consequently, the Hall conductivity must satisfy $\sigma_{xy}(-B) = -\sigma_{xy}(B)$. Therefore, $\sigma_{xy}$ is odd in $B$ and vanishes at $B=0$. For small $B$, $\sigma_{xy}$ is expected to be linear in $B$, or equivalently in the scalar spin chirality.  

\par Other symmetry properties of $\hat{\mathcal{H}}_{\boldsymbol{k}}$ are considered in Supplementary, Sec.~D.

\subsubsection{\label{sec:orb} Orbital magnetization}
\par The emergence of the anomalous Hall effect suggests the presence of an underlying order parameter associated with broken time-reversal symmetry. In our previous work, we identified this order parameter as the orbital magnetization $\mathcal{M}$, which closely follows the behavior of the AHE~\cite{npjQM2026}. In nearly collinear centrosymmetric antiferromagnets, $\mathcal{M}$ is directly related to the Berry curvature and therefore mirrors the AHE~\cite{PRB2026}. Below we show that an analogous mechanism operates in trimerized kagome lattices. The crucial difference from the itinerant mechanism discussed in Refs.~\cite{npjQM2026,PRB2026} is the presence of a strong on-site contribution to $\mathcal{M}$, originating from the imbalance between the occupations of the $k=0$ and $k=2$ molecular-trimer states.

\par The orbital magnetization (per trimer) is given by the BZ integral:
\noindent
\begin{displaymath}
\mathcal{M} = \frac{\sqrt{3}\mu}{2} \sum_{\nu = \pm} \int_{\rm BZ} \frac{d \boldsymbol{k}}{(2 \pi)^{2}} f^{\nu}_{\boldsymbol{k}}  \mathcal{M}_{\boldsymbol{k}}^{\nu},
\end{displaymath}
\noindent where $\frac{\sqrt{3}}{2}$ is the unit-cell area and the parameter $\mu$ was introduced in Sec.~\emph{Single trimer}. The momentum-resolved orbital magnetization is
\noindent
\begin{equation}
\mathcal{M}_{\boldsymbol{k}}^{-} = -\mathcal{M}_{\boldsymbol{k}}^{+} = \cos 2 \theta_{\boldsymbol{k}} = -\frac{\mathrm{h}^{z}_{\boldsymbol{k}}+B}{|\mathbf{h}^{\phantom{z}}_{\boldsymbol{k}}+\boldsymbol{B}|}.
\label{eq:Morb}
\end{equation}
\noindent 
\par Using Eq.~(\ref{eq:TRinv}), it follows that $\mathcal{M}_{-\boldsymbol{k}}^{\nu}(B) = -\mathcal{M}_{\boldsymbol{k}}^{\nu}(-B)$, in complete analogy with the transformation of the Berry curvature. Consequently, the total orbital magnetization is an odd function of $B$, $\mathcal{M}(B) = - \mathcal{M}(-B)$, and therefore vanishes in the coplanar limit $B=0$. Thus, just as the anomalous Hall conductivity, the orbital magnetization emerges only when the noncoplanar spin texture breaks time-reversal symmetry.

\subsubsection{\label{sec:ppoints} Points of phase diagram}
\par
\begin{figure}[b]
\begin{center}
\includegraphics[width=8.6cm]{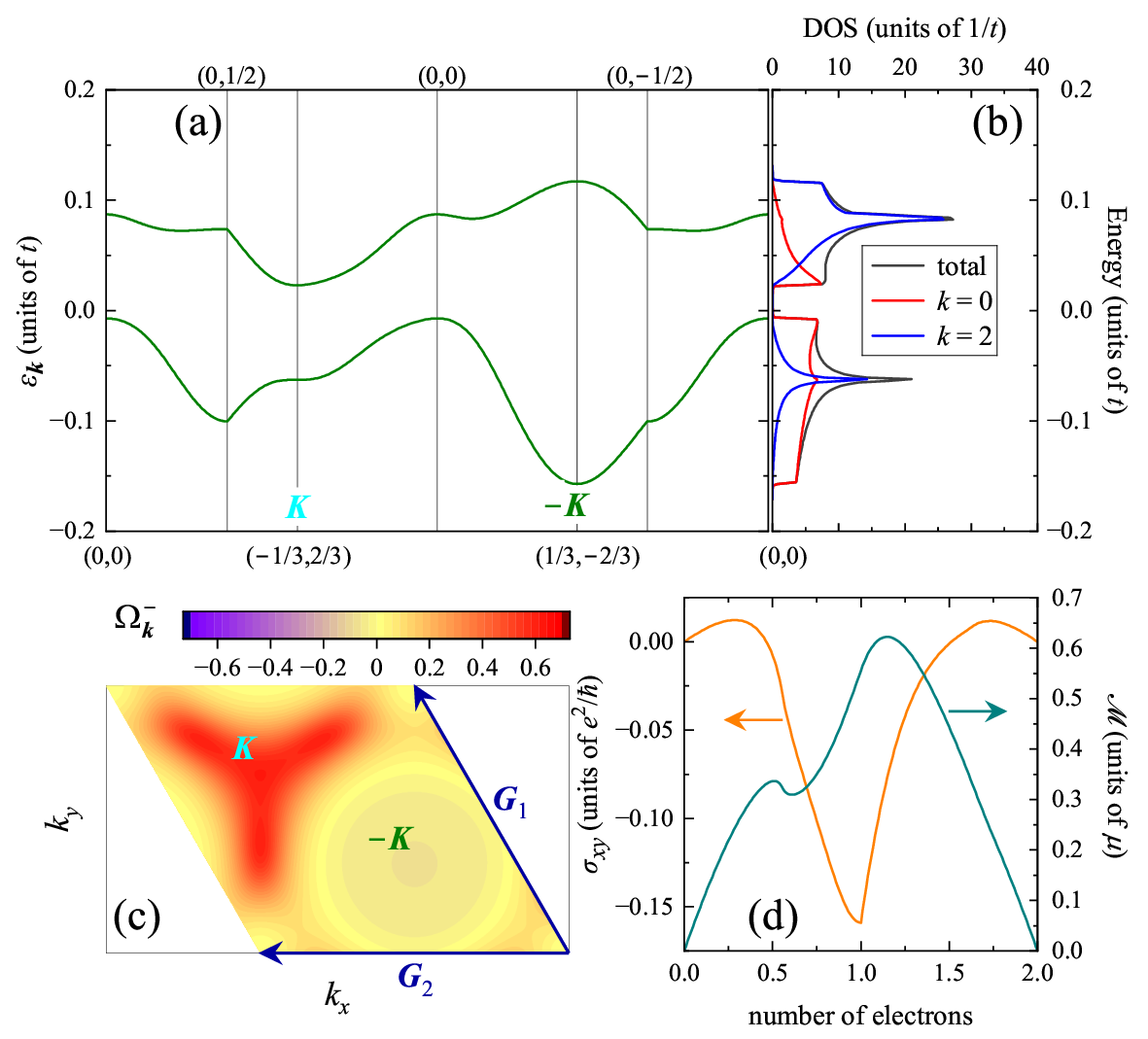} 
\end{center}
\caption{Summary of the properties in the Chern-insulator regime with $t'=0.1t$, $t_{2}=-0.01t$, and $\vartheta = 0.51\pi$. (a) Band structures along high-symmetry directions. (b) Total and partial densities of states (DOS). (c) Berry-curvature contour. (d) Band filling dependence of the Hall conductivity $\sigma_{xy}$ and the orbital magnetization $\mathcal{M}$. High-symmetry points are given in units of the reciprocal-lattice vectors $\boldsymbol{G}_{1}$ and $\boldsymbol{G}_{2}$. The locations of $\boldsymbol{K}$ and $-\boldsymbol{K}$ are indicated for reference.}
\label{fig:CI}
\end{figure}

\begin{figure}[t]
\begin{center}
\includegraphics[width=8.6cm]{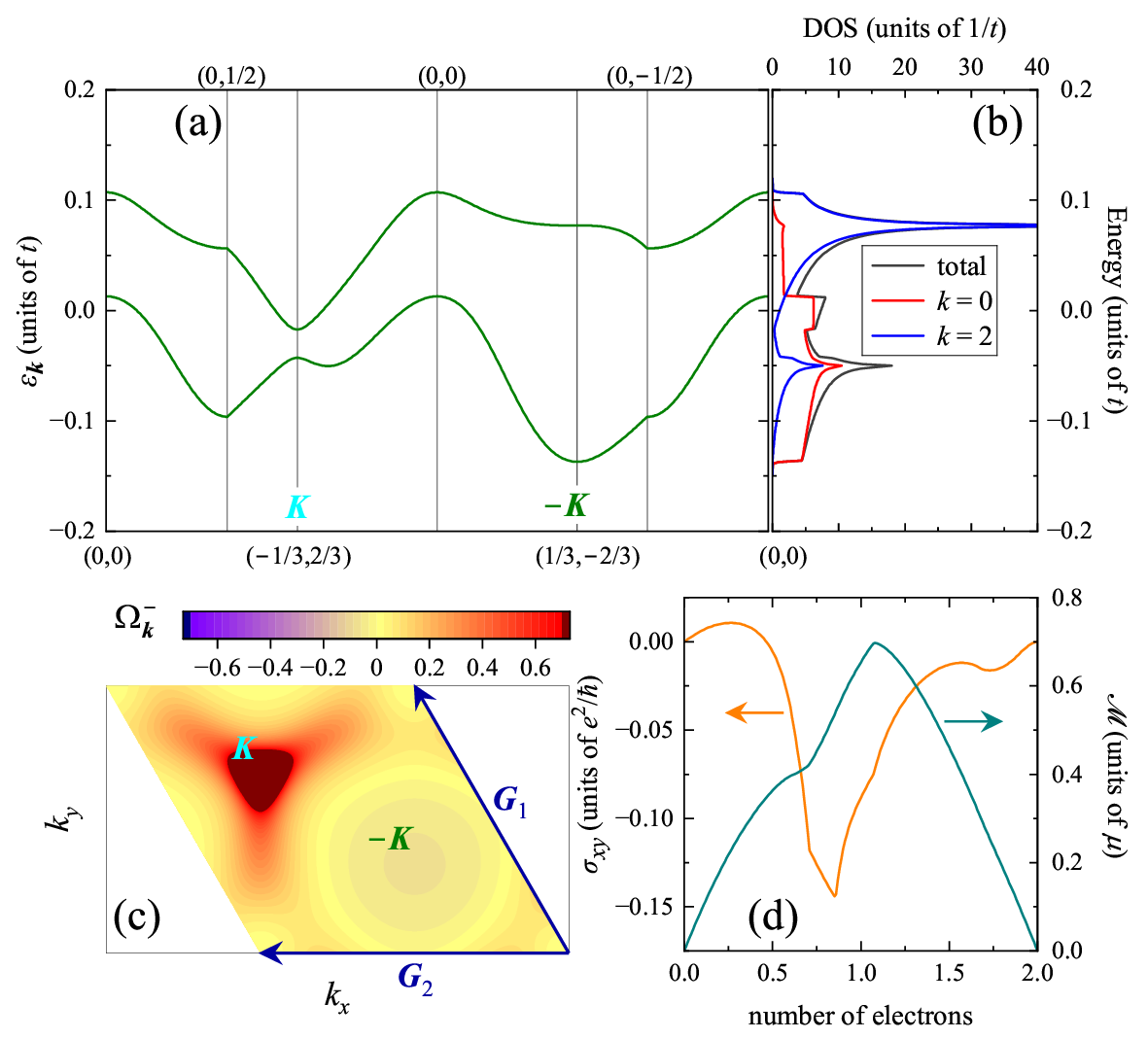} 
\end{center}
\caption{Summary of the properties in the Chern-semimetal regime with $t'=0.1t$, $t_{2}=0.01t$, and $\vartheta = 0.51\pi$. See Fig~\ref{fig:CI} for details.}
\label{fig:CM}
\end{figure}

\begin{figure}[b]
\begin{center}
\includegraphics[width=8.6cm]{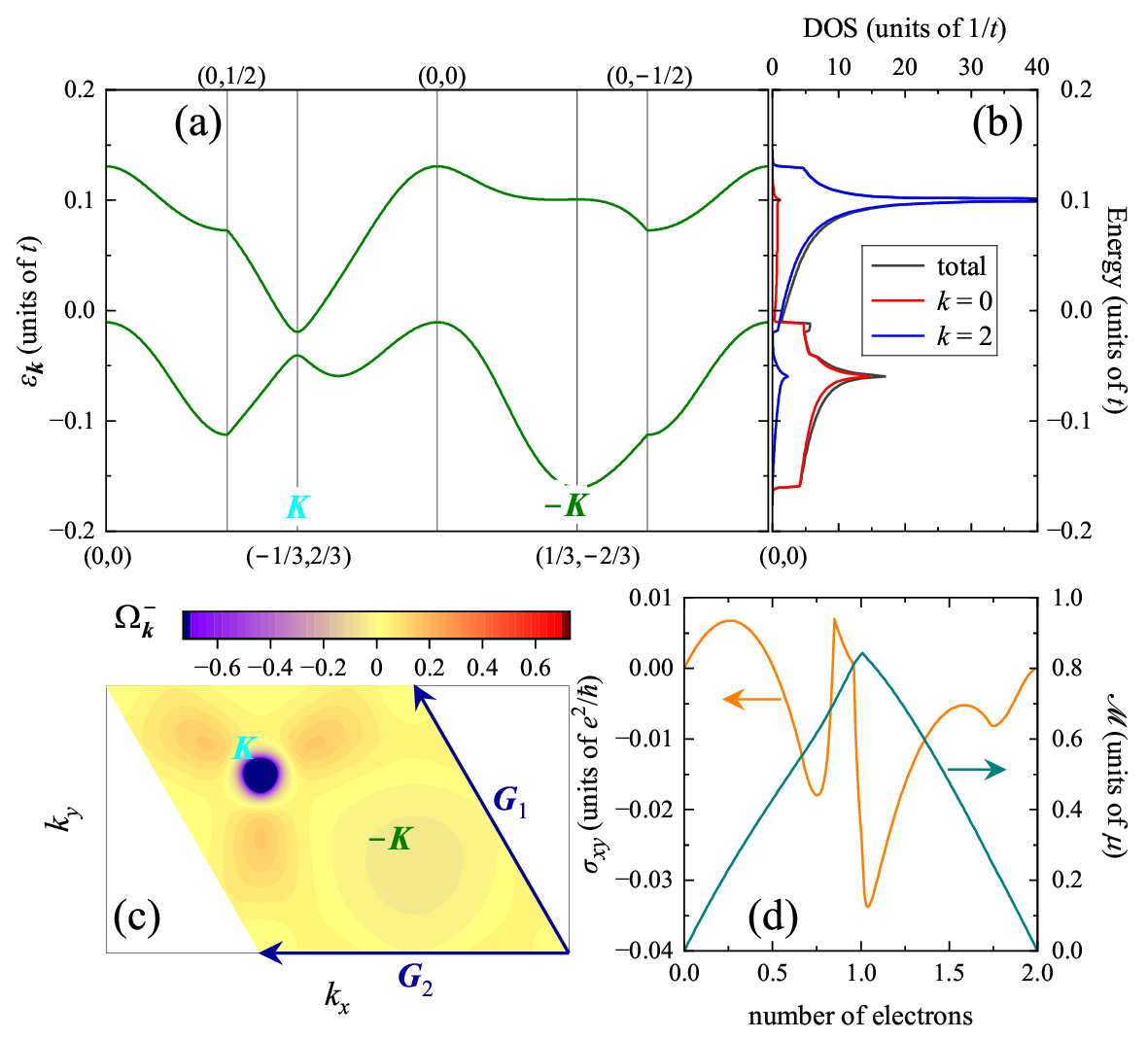} 
\end{center}
\caption{Summary of the properties in the normal-semimetal regime with $t'=0.1t$, $t_{2}=0.01t$, and $\vartheta = 0.515\pi$. See Fig~\ref{fig:CI} for details.}
\label{fig:NM}
\end{figure}

\begin{figure}[t]
\begin{center}
\includegraphics[width=8.6cm]{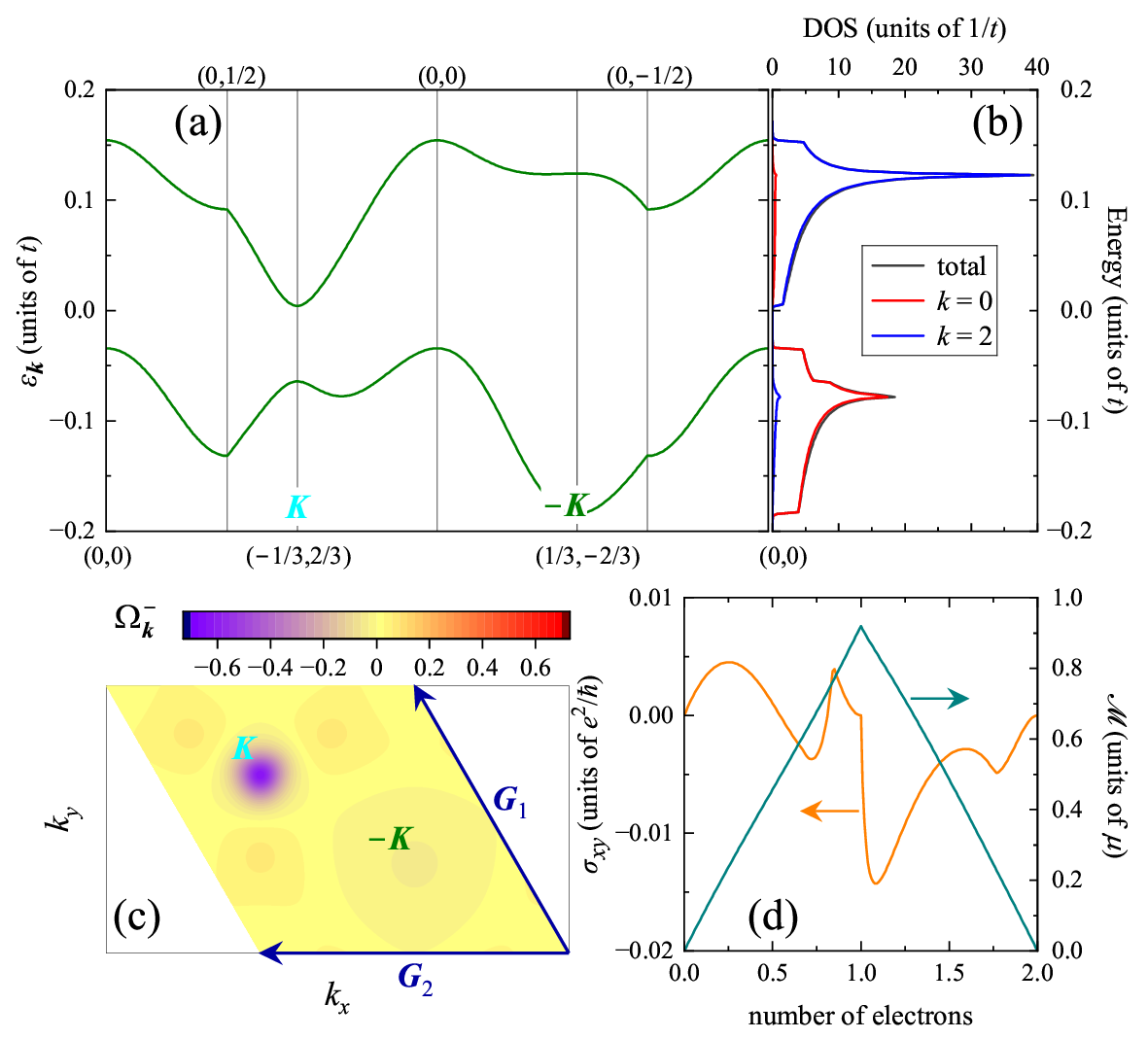} 
\end{center}
\caption{Summary of the properties in the normal-insulator regime with $t'=0.1t$, $t_{2}=0.01t$, and $\vartheta = 0.52\pi$. See Fig~\ref{fig:CI} for details.}
\label{fig:NI}
\end{figure}

 In this section we briefly discuss four representative points of the phase diagram, each illustrating a distinct regime: $(t',t_{2},\vartheta) = (0.1t,-0.01t,0.51\pi)$ (Chern insulator with $\mathcal{C}=1$), $(0.1t,0.01t,0.51\pi)$ (Chern semimetal in the compensated, indirect-gap sense), $(0.1t,0.01t,0.515\pi)$ (normal compensated semimetal), and $(0.1t,0.01t,0.52\pi)$ (normal insulator); see Fig.~\ref{fig:PhaseD}. The corresponding band structures, densities of states, Berry curvatures, and band-filling dependencies of $\sigma_{xy}$ and $\mathcal{M}$ are shown in Figs.~\ref{fig:CI}-\ref{fig:NI}.  
\noindent
\noindent
\noindent 
\noindent

\par In the Chern-insulator regime, the dominant positive contribution to $\Omega_{\boldsymbol{k}}^{-}$ near $\boldsymbol{K}$ outweighs the negative contribution near $-\boldsymbol{K}$, yielding $\mathcal{C}=1$ and quantized Hall conductivity $\sigma_{xy} = \frac{1}{2\pi}$ at half-filling (Fig.~\ref{fig:CI}). In our convention, $\boldsymbol{K} = \frac{2}{3} \left( \boldsymbol{G}_{1} + \boldsymbol{G}_{2} \right)$, where the reciprocal-lattice vectors are $\boldsymbol{G}_{1}=2\pi(-\frac{1}{\sqrt{3}},1)$ and $\boldsymbol{G}_{2}=2\pi(-\frac{2}{\sqrt{3}},0)$. 

\par Upon moving into the Chern-semimetal regime, the direct band gap at $\boldsymbol{K}$ decreases, and the indirect gap between the states $\varepsilon_{\boldsymbol{\Gamma}}^{-}$ and $\varepsilon_{\boldsymbol{K}}^{+}$ closes (Fig.~\ref{fig:CM}), where $\boldsymbol{\Gamma}=(0,0)$. This leads to a redistribution of $\Omega_{\boldsymbol{k}}^{-}$, with an enhanced weight at $\boldsymbol{K}$ itself and a corresponding reduction in the surrounding regions. As a result, the Chern number remains $\mathcal{C}=1$, but due to the finite band overlap between $\varepsilon_{\boldsymbol{k}}^{-}$ and $\varepsilon_{\boldsymbol{k}}^{+}$, the Hall conductivity $\sigma_{xy}$ at half-filling deviates from its quantized value $\frac{1}{2\pi}$.

\par Across the transition $\mathcal{C}=1 \to \mathcal{C}=0$, which is accompanied by a gap closing at $\boldsymbol{K}$ and its reopening on the other side of the transition, $\Omega_{\boldsymbol{k}}^{-}$ undergoes a singular redistribution. Such a redistribution is realized in the normal-semimetal phase, where $\Omega_{\boldsymbol{k}}^{-}$ is characterized by a large \emph{negative} peak at $\boldsymbol{K}$, compensated by three positive regions surrounding $\boldsymbol{K}$ (Fig.~\ref{fig:NM}). Consequently, $\mathcal{C}$ drops from $1$ to $0$, and the corresponding Hall conductivity is strongly suppressed. Nevertheless, at half-filling $\sigma_{xy}$ remains finite due to the incomplete cancellation of the contributions from the $\nu = \pm$ states. 

\par With increasing $\vartheta$, the splitting between the $\nu = \pm$ bands grows, and the system eventually becomes insulating. As the direct gap at $\boldsymbol{K}$ increases, the corresponding negative peak of $\Omega_{\boldsymbol{k}}^{-}$ diminishes, as do the three positive contributions surrounding $\boldsymbol{K}$ (Fig.~\ref{fig:NI}). The Chern number remains $\mathcal{C}=0$. Furthermore, because the occupied band carries $\mathcal{C}=0$, the reopening of the gap yields $\sigma_{xy}=0$ at half-filling.

\par In all considered examples, a finite spin chirality ($\vartheta \ne \frac{\pi}{2}$) generates an effective field $B$ that splits the $k=0$ and $k=2$ molecular-trimer states. Because these states carry opposite values of $\langle k | \hat{\mathcal{L}}^{z} | k \rangle$, the resulting population imbalance produces a finite orbital magnetization $\mathcal{M}$, whose magnitude depends on the band filling.

\subsubsection{\label{sec:flux} Time-reversal symmetry breaking without Zeeman field}
\par In this section we explore a hypothetical mechanism for breaking time-reversal symmetry in the absence of $B$. Our aim is to identify a hopping pattern that violates Eq. (\ref{eq:TRinv}) while preserving the threefold rotational symmetry. Whether such a hopping pattern can be realized physically remains an open question.

\par First, let us consider a formal construction. A uniform shift $\boldsymbol{k} \to \boldsymbol{k}+\boldsymbol{q}$ of all Bloch wave vectors is equivalent to a gauge transformation of the hopping amplitudes and therefore leaves unchanged any quantity obtained from Brillouin-zone integrals. The idea is then to apply distinct shifts to the nn and nnn inter-trimer hoppings in the original kagome setting. For instance, shifting the $\boldsymbol{k}$-mesh for the nn hoppings by $- \boldsymbol{K}$ results in the replacement $\gamma_{i}^{0} \to \gamma_{i}^{+}$ in Eqs.~(\ref{eq:hk1})-(\ref{eq:hk4}), whereas shifting the mesh for the nnn hoppings by $\boldsymbol{K}$ yields the replacement $\gamma_{i}^{0} \to \gamma_{i}^{-}$ in Eqs.~(\ref{eq:dhk1})-(\ref{eq:dhk4}). Thus, although the nn and nnn contributions separately satisfy Eq.~(\ref{eq:TRinv}) up to gauge transformations, no single global gauge transformation can restore Eq. (\ref{eq:TRinv}) for their sum. Consequently, the total Hamiltonian violates time-reversal symmetry. Nevertheless, because threefold rotations map $\boldsymbol{K}$ onto itself modulo reciprocal lattice vectors, $\mathcal{C}_{3z}^{+}\boldsymbol{K} = \boldsymbol{K} + \boldsymbol{G}_{1}$ and $\mathcal{C}_{3z}^{-}\boldsymbol{K} = \boldsymbol{K} - \boldsymbol{G}_{2}$, the threefold rotational symmetry remains intact.

\par
\begin{figure}[h]
\begin{center}
\includegraphics[width=8.6cm]{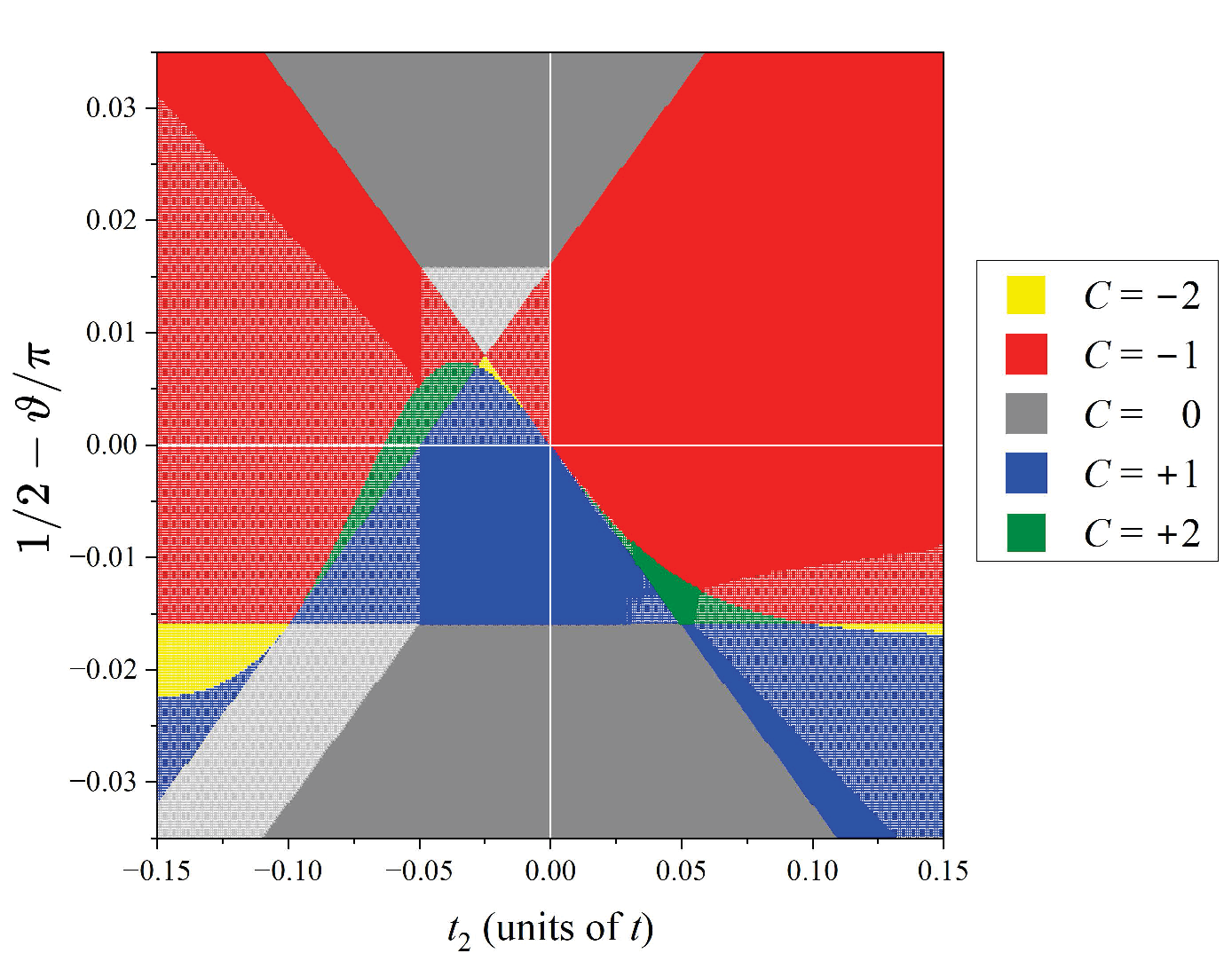} 
\end{center}
\caption{Phase diagram at half filling for the modified model as a function of the next-nearest-neighbor hopping $t_{2}$ and the polar angle $\vartheta$, shown in the vicinity of the coplanar configuration $\vartheta=\frac{\pi}{2}$. Further details are provided in Fig~\ref{fig:PhaseD}.}
\label{fig:PhaseDflux}
\end{figure}

 The corresponding phase diagram is rather rich (Fig.~\ref{fig:PhaseDflux}). In addition to the phases with $\mathcal{C}=0$ and $\pm 1$, it also contains the phase with $\mathcal{C}=\pm 2$. Moreover, even in the coplanar case for $\vartheta=\frac{\pi}{2}$ several topological phases with $\mathcal{C}=\pm 1$ and $\mathcal{C}=2$ can occur. In particular, starting from $t_{2}=0$ and $\vartheta=\frac{\pi}{2}$, the phase with $\mathcal{C}=1$ can be reached either by introducing negative spin chirality ($\vartheta > \frac{\pi}{2}$) or by turning on a negative nnn hopping $t_{2}$. Likewise, the phase with $\mathcal{C}=-1$ can be obtained either for $\vartheta < \frac{\pi}{2}$ or for positive $t_{2}$.
\noindent

\par The above construction may appear somewhat formal. However, it admits an equivalent interpretation in terms of Peierls phases attached to the bonds, closely analogous to Haldane's original model, where complex hopping phases mimic the effect of a periodic magnetic field with zero net flux~\cite{Haldane}. Specifically, we leave the intra-trimer hoppings unchanged and modify the nn and nnn inter-trimer hoppings as $t' \to t'e^{\pm i\phi_{\textrm{nn}}}$ and $t_{2} \to t_{2}e^{\pm i\phi_{\textrm{nnn}}}$, where the `$+$' and `$-$' signs correspond to hoppings between the atomic sites along the directions $1 \to 2 \to 3 \to 1$ and $1 \to 3 \to 2 \to 1$, respectively. Threefold rotational invariance requires that the phases   $\phi_{\textrm{nn}}$ and $\phi_{\textrm{nnn}}$ take the same values on all nn and nnn bonds. For a periodic magnetic field, the total flux through the unit cell must be $0$ ($\textrm{mod}~2\pi$, in units of $\frac{e}{\hbar}$). Since each triangular loop contains three equivalent bonds, the condition of trivial total flux through the loop requires $\phi_{\textrm{nn}}=\pm \frac{2\pi}{3}$ and $\phi_{\textrm{nnn}}=\pm \frac{2\pi}{3}$, i.e., the total flux through each black and blue dashed-line triangle in Fig.~\ref{fig:kagome} should take the trivial value $2\pi$. Obviously, the choice $\phi_{\textrm{nn}}=\phi_{\textrm{nnn}}$ corresponds to a global gauge transformation and therefore does not change the symmetry. On the contrary, the choice $\phi_{\textrm{nn}}=-\phi_{\textrm{nnn}}$ breaks the time-reversal symmetry. Since $\boldsymbol{K} \cdot \boldsymbol{a}_{i} = -\frac{2\pi}{3}~(\textrm{mod}~2\pi)$ for $i=1$, $2$, and $3$, the above $\boldsymbol{k}$-mesh shift by $\pm \boldsymbol{K}$ for the nn and nnn hoppings is fully equivalent to the phase choice $\phi_{\textrm{nn}}=-\phi_{\textrm{nnn}}$.

\par Therefore, the relation $\mathrm{h}^{z}_{-\boldsymbol{k}} \ne -\mathrm{h}^{z}_{\boldsymbol{k}}$ is no longer satisfied. Consequently, Eq.~(\ref{eq:Morb}) no longer enforces the cancellation of opposite-$\boldsymbol{k}$ contributions, and a finite orbital magnetization may arise even when $B=0$.

\section*{DISCUSSION}
\par Historically, the study of time-reversal-symmetry breaking in antiferromagnets can be traced to the phenomenon of weak ferromagnetism, discovered experimentally in 1957~\cite{BorovikRomanov} and explained theoretically shortly thereafter~\cite{Dzyaloshinskii_weakF}. We have argued that a broad class of time-reversal-odd phenomena including the AHE in nearly collinear and noncollinear antiferromagnets can be understood within this common framework. The central ingredient is the orbital magnetization generated by the SO coupling and/or the scalar spin chirality, which provides a direct link between weak ferromagnetism and the AHE.

\par In centrosymmetric antiferromagnets, the net spin magnetization may vanish owing to a cancellation of the spin canting induced by the DM interactions, as occurs, for example, in RuO$_2$~\cite{SmejkalSA}. More generally, the SO-interaction component responsible for weak spin ferromagnetism, which has the same sign on all bonds connected to a given magnetic site, can be gauged away by a local rotation of the spin reference frame~\cite{PRB2026,npjQM2026}. In this sense, the presence or absence of a net spin magnetization is not essential for the AHE. Another important point is that both the AHE and the net orbital magnetization are driven by the same sign-alternating component of the SO interaction that does not contribute to the ferromagnetic canting of spins~\cite{PRB2026,npjQM2026}. Thus, AHE in centrosymmetric nearly collinear antiferromagnets can be viewed as a manifestation of weak orbital ferromagnetism. This interpretation is fully consistent with the original phenomenological theories of weak ferromagnetism, which are formulated in terms of the total magnetic moment rather than its separate spin and orbital contributions, since both transform identically under the crystal symmetry operations~\cite{TurovBook,TurovUFN}. In contrast, microscopic theories of DM interactions are usually formulated in terms of spin-only moments~\cite{Moriya_weakF,review2024} and therefore cannot directly address the problem of weak orbital ferromagnetism. Extending these theories to explicitly incorporate orbital magnetism is considerably more challenging, although several attempts have been made~\cite{Katsnelson}.

\par The same physical picture emerges in noncollinear kagome antiferromagnets. Here, spin chirality and SO coupling can be reformulated in terms of an effective Zeeman field that lifts the degeneracy between molecular-orbital states with opposite angular momenta $m= \pm 1$. Alternatively, net orbital magnetization may arise from a gauge field generated by complex inter-trimer electron hoppings, even in the absence of an effective Zeeman field. Both mechanisms ultimately lead to weak orbital ferromagnetism accompanied by the AHE. 

\par A unique aspect of noncollinear antiferromagnetism on the trimerized kagome lattice is that, in addition to the conventional AHE, it provides a natural platform for the realization of the quantum anomalous Hall effect (QAHE)~\cite{Ohgushi,SCZhang2}. Remarkably, the corresponding QAHE phases with $C= \pm 1$ may be induced solely by scalar spin chirality and do not, in principle, require SO coupling. This mechanism differs fundamentally from those proposed for centrosymmetric antiferromagnets, where the realization of topological phases typically relies on additional ingredients such as Rashba spin-orbit coupling and staggered electrostatic potentials~\cite{NakaqAHE}. Moreover, the pure emergent-gauge-field scenario gives rise to an even richer phase diagram: in addition to the $C= \pm 1$ phases, it also stabilizes topological phases with $C= \pm 2$. Thus, the trimerized kagome antiferromagnet illustrates how the concept of weak orbital ferromagnetism, rooted in the physics of the 1950s, naturally extends to encompass a rich landscape of topological phenomena. We suggest that the weak net magnetic moment experimentally observed in Mn$_3$Ge and related kagome materials~\cite{Nakatsuji2} may be predominantly orbital in origin.

\section*{Methods}
\subsection{Details of calculations}
\par The Chern numbers were evaluated using the Fukui-Hatsugai-Suzuki method~\cite{FHS} on a uniform $301 \times 301$ $\boldsymbol{k}$-point mesh covering the first Brillouin zone. The anomalous Hall conductivity, orbital magnetization, and density of states were calculated on a uniform $5901 \times 5901$ $\boldsymbol{k}$-point mesh covering the first Brillouin zone. 

\section*{Data availability}
\par The data supporting the findings of this study are available within this article. Additional data are available from the corresponding author on reasonable request.

\section*{Acknowledgement}
\par I am grateful to Akihiro Tanaka, Ryotaro Arita, and Shinichiro Seki for valuable discussions. MANA is supported by World Premier International Research Center Initiative (WPI), MEXT, Japan. This research did not receive finding.

\section*{Author contributions}
\par I.V.S. conceptualized the work, performed the calculations, and wrote the manuscript. 

\section*{Competing interests}
\par The author declare no competing financial or non-financial interests.

\section*{Additional information}
\par \textbf{Supplementary information} The online version contains supplementary material available at ...

\par \textbf{Correspondence} and requests for materials should be addressed to I. V. Solovyev.

\end{document}